\documentclass[final,3p,twocolumn,times,hidelinks]{elsarticle}

\usepackage{natbib}
\usepackage[hyphens]{url}
\usepackage{hyperref}

\usepackage{graphics}
\usepackage{graphicx}

\usepackage{amssymb}

\newcounter{bla}

\def\figwidth{3.083in}

\usepackage{amsmath}
\usepackage{xcolor}
\usepackage{graphicx}
\usepackage{amssymb}
\usepackage{physics}

\usepackage{multirow}

\usepackage[color=green!60]{todonotes}

\newcommand{\Eqref}[1]{Eq.~\eqref{#1}}
\newcommand{\Figref}[1]{Fig.~\ref{#1}}

\journal{Computer Physics Communications}

\begin{document}

\begin{frontmatter}



\title{Reduction of Autocorrelation Times in Lattice Path Integral Quantum Monte Carlo via Direct Sampling of the Truncated Exponential Distribution}


\author[a,b]{Emanuel Casiano-Diaz\corref{author}}
\author[c]{Kipton Barros}
\author[b]{Ying Wai Li}
\author[a,d,e]{Adrian Del Maestro}

\cortext[author] {Corresponding author.\\\textit{E-mail address:} ecasiano@vols.utk.edu}
\address[a]{Department of Physics and Astronomy, University of Tennessee, Knoxville, TN 37996, USA}
\address[b]{Computer, Computational, and Statistical Sciences Division, Los Alamos National Laboratory, Los Alamos, NM 87545, USA}
\address[c]{Theoretical Division and Center for Nonlinear Studies, Los Alamos National Laboratory, Los Alamos, NM 87545, USA}
\address[d]{Min H. Kao Department of Electrical Engineering and Computer Science, University of
Tennessee, Knoxville, TN 37996, USA}
\address[e]{Institute for Advanced Materials and Manufacturing, University of Tennessee, Knoxville,TN 37996, USA}

\begin{abstract}
    In Monte Carlo simulations, proposed configurations are accepted or rejected according to an acceptance ratio, which depends on an underlying probability distribution and an \emph{a priori} sampling probability. By carefully selecting the probability distribution from which random variates are sampled, simulations can be made more efficient, by virtue of an autocorrelation time reduction. In this paper, we illustrate how to directly sample random variates from a two dimensional truncated exponential distribution. 
We show that our direct sampling approach converges faster to the target distribution compared to rejection sampling. 
The direct sampling of one and two dimensional truncated exponential distributions is then applied to a recent Path Integral Monte Carlo (PIMC) algorithm for the simulation of Bose-Hubbard lattice models at zero temperature. The new sampling method results in improved acceptance ratios and reduced autocorrelation times of estimators, providing an effective speed up of the simulation.
\end{abstract}


\begin{keyword}
Autocorrelation Time; Monte Carlo; Truncated Exponential Distribution; Direct Sampling.

\end{keyword}

\end{frontmatter}



\section{Introduction}
\label{sec:introduction}

Sequential samples obtained in the random walk of a Markov Chain Monte Carlo (MCMC) simulation generally exhibit statistical correlations. The quality of a statistical estimate is directly related to the number of effectively uncorrelated samples obtained. A key challenge in the development of MCMC methods is therefore the reduction of computational time required to generate well-decorrelated samples.

One of the most ubiquitous MCMC methods is the Metropolis Algorithm~\cite{doi:10.1063/1.1699114,binder_heermann_montecarlo,krauth06,Giordano:2006aa}, where samples are obtained from a probability distribution that is often non-trivial to sample. In this algorithm, the principle of detailed balance leads to a non-negative acceptance ratio, $R$, for determining if randomly proposed configurations are accepted or rejected. Proposed configurations are only kept when this acceptance ratio is larger than a random number drawn from the uniform distribution, $r \sim \mathcal{U} \left ( 0, 1 \right )$, such that $r < R$, and rejected otherwise. In most applications it is common to encounter cases in which the acceptance ratio is small, leading to an inefficient Markov Chain as most proposed configurations are rejected. By carefully choosing the underlying probability distribution from which random variates in a Monte Carlo update are sampled, the acceptance ratio can be increased and even become unity so that every new configuration is accepted, thus improving the dynamics of the random walk and decreasing the correlation between subsequent samples.

In a recently developed Path Integral Monte Carlo (PIMC) algorithm~\cite{CasianoDiaz:2022pf}, the acceptance ratio of certain updates depends on drawing random variates from a one or two dimensional truncated exponential distribution -- an exponential distribution restricted to a finite domain. In this paper, we describe how to directly sample two random variates from a two dimensional truncated exponential distribution, and apply this sampling strategy in PIMC. The resulting method leads to a reduction of autocorrelation times and therefore faster convergence of statistical averages to their exact values.

The paper is organized as follows: In Section~\ref{sec:review1D} we review how to sample variates from a one dimensional probability distribution by inverting the cumulative distribution function (CDF) of a probability density function (PDF). We will do this in the context of the one dimensional truncated exponential distribution. In Section~\ref{sec:sampling2D}, this method is then generalized to the non-trivial case of directly sampling two random variates from a two dimensional truncated exponential distribution. The direct sampling of variates from both one and two dimensional truncated exponential distributions is then applied to the QMC algorithm of Ref.~\cite{CasianoDiaz:2022pf} for the simulation of bosonic lattices at zero temperature and it is shown that direct sampling results in decreased autocorrelation times for the kinetic and potential ground state energies at no performance cost.

\section{Direct sampling of $1D$ truncated exponential distribution}
\label{sec:review1D}

The one dimensional $(1D)$ truncated exponential distribution is defined as:
\begin{equation}
P_1(x) = \frac{1}{\mathcal{Z}} e^{-c(x-a)} = \frac{c e^{-c(x-a)}}{1-e^{-c(b-a)}},
\label{eq:truncexpon_simple}
\end{equation}
where $a$ and $b$ are the lower and upper bounds of the finite domain, respectively, $c$ is a scale parameter, and $x$ is a random variable in the truncation interval satisfying $a \leq x \leq b$. The factor $\mathcal{Z}$
has been chosen to ensure that the distribution is normalized: $\int_{a}^{b} \dd{x} P_1(x)=1$.
The cumulative distribution function (CDF) of $P_1(x)$ is, 
\begin{equation}
    F_1(x) \equiv \int_{a}^x \dd{x^\prime} P_1(x^\prime)  = \frac{1-e^{-c(x-a)}}{c \mathcal{Z}}\, .
\label{eq:truncexpon_simple_cdf}
\end{equation}

The inverse transform sampling method inverts the functional dependence $y = F_1(x)$ to obtain samples from the target distribution  $x \sim P_1(x)$ of~\Eqref{eq:truncexpon_simple}. The first step is to sample a random variable $y$ uniformly between 0 and 1. We denote this random variable $y \sim U(0,1)$.
Then $x = F_1^{-1}(y)$ yields a random variable with the desired target distribution, $x \sim P_1(x)$.
Inverting the CDF in \Eqref{eq:truncexpon_simple_cdf} we find:
\begin{equation}
x(y) = a - \frac{\ln \left ( 1 - c\mathcal{Z}y \right )}{c}\, .
\label{eq:truncexpon_simple_rvs}
\end{equation}

When the CDF cannot be analytically inverted, a common practical approach attributed to von Neumann is rejection sampling~\cite{forsytheNeumannComparisonMethod1972,newmanb99}, which allows for the brute force sampling of $P_1(x)$ on a finite domain with $P_{1,\rm max} \equiv \max_{a \le x \le b} P_1(x)$ through the sequential comparison of two independently sampled random numbers.

\begin{center}
\fcolorbox{black}{white}{\parbox{\columnwidth}{
\centering{\textbf{Rejection Sampling}}
\begin{enumerate}
    \item Sample a random number from the uniform distribution $x \sim U(a,b)$.
    \item Sample independently another random number from the uniform distribution $\chi \sim U(0,P_{\rm 1, max})$.
    \item If $\chi < P_1(x)$ then accept $x$. Otherwise, reject the proposal and return to step 1.
\end{enumerate}}}
\end{center}
The value $x$ returned by this procedure will be a good sample from the distribution $P_1$. Note, however, that if $P_1$ deviates strongly from $U(a,b)$ then there are likely to be many rejections before a good sample is returned.

\begin{figure}[t]
\begin{center}
    \includegraphics[width=\figwidth]{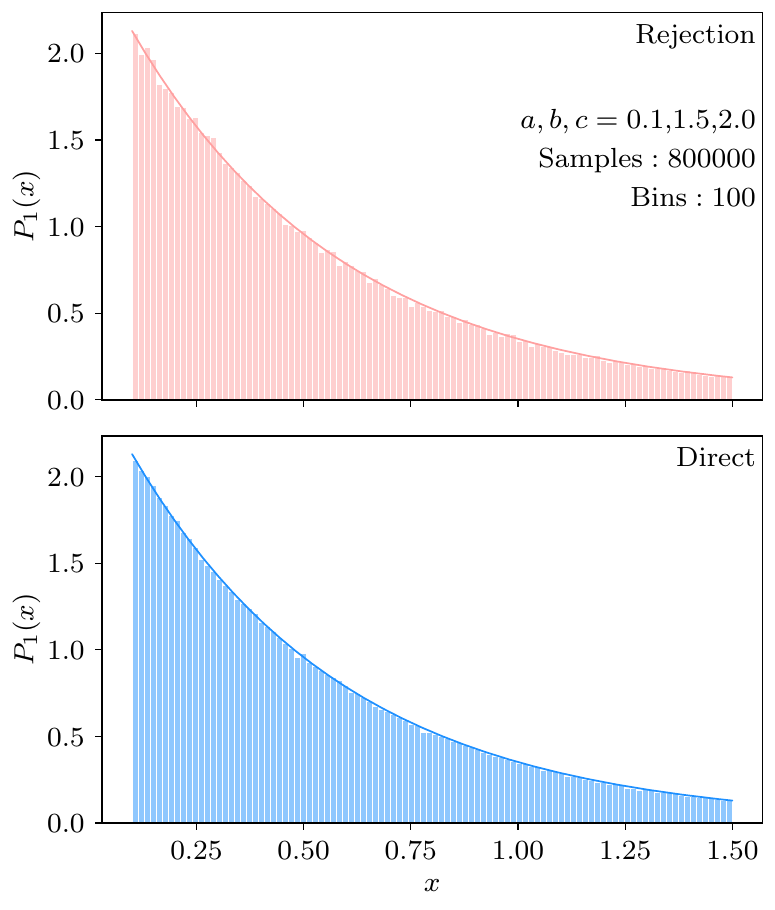}
\end{center}
\caption{One dimensional truncated exponential distribution \Eqref{eq:truncexpon_simple} generated via rejection sampling (top) and direct sampling (bottom). For the number of samples shown ($800,000$), both methods successfully generate the desired one dimensional truncated exponential distribution of the random variate $x$. The direct sampling data is closer to the exact distribution (solid curve) because it includes more good samples.} 
\label{fig:simple_truncexpon_histogram_benchmark}
\end{figure}

While rejection sampling is not necessary here due to the existence of the inverse, to setup our analysis of the $2D$ case, we compare direct and rejection sampling by generating a histogram of random samples $x \sim P_1$ using both methods.  The results are shown in \Figref{fig:simple_truncexpon_histogram_benchmark} for $8\times10^5$ samples.  Both histograms agree with the expected result in \Eqref{eq:truncexpon_simple}, but fluctuations are larger for the rejection sampling case.  
This can be attributed to the large number of good samples in the direct sampling dataset, since random variates obtained from \Eqref{eq:truncexpon_simple} are always accepted. In contrast, a significant fraction of iterations for the rejection sampling method did not lead to a good sample. 

\begin{figure}[t]
\begin{center}
    \includegraphics[width=\figwidth]{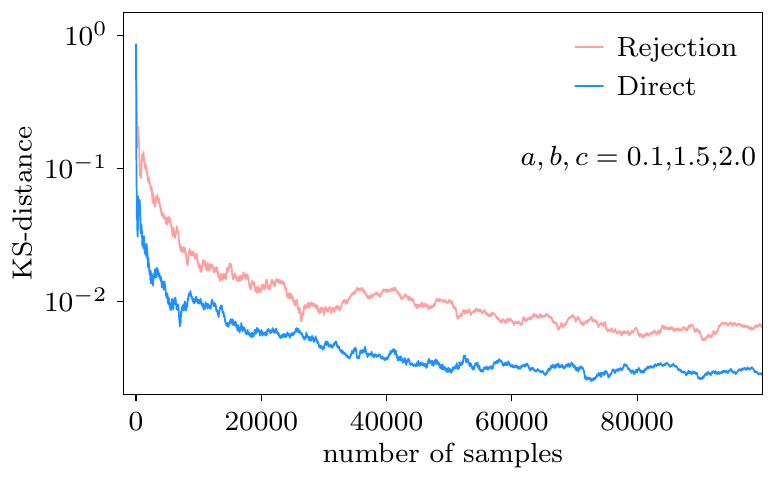}
\end{center}
\caption{Kolmogorov-Smirnov (KS) test results. The KS-distance is plotted as a function of number of samples for two sampling methods. The KS-distance is smaller for the direct sampling dataset, which indicates that it better estimates the true one dimensional truncated exponential distribution, \Eqref{eq:truncexpon_simple}.} 
\label{fig:ks_test}
\end{figure}

For a one dimensional probability distribution, such as $P_1(x)$, a Kolmogorov-Smirnov (KS) test can be used to quantify how well the random variate dataset follows the target distribution. \Figref{fig:ks_test} shows the KS-distance or KS-statistic, which measures the maximum difference between empirical and theoretical CDFs, as a function of the number of samples in the dataset. 
For the direct sampling dataset, the KS-distance decays much faster than for the rejection sampling dataset. This is expected, because the rejection sampling dataset contains fewer statistical samples from~\Eqref{eq:truncexpon_simple}.

Having reviewed the standard methods of direct inversion and rejection sampling for the $1D$ truncated exponential distribution, we now generalize to the $2D$ case which is relevant for Path Integral quantum Monte Carlo simulations. 

\section{Direct sampling of $2D$ truncated exponential distribution}
\label{sec:sampling2D}

The probability distribution \Eqref{eq:truncexpon_simple} can be generalized to two random variables $x_1,x_2$ as:
\begin{equation}
P_2(x_1,x_2) = \frac{1}{\mathcal{Z}_J} e^{-c(x_2-x_1)},
\label{eq:truncexpon_joint}
\end{equation}
where $a \leq x_1 < x_2 \leq b$ and the distribution is normalized by:
\begin{equation}
\mathcal{Z}_J = \frac{e^{-c \left ( b-a \right )} - ac + bc - 1}{c^2} .
\end{equation}
The random variables $x_1$ and $x_2$ can be sampled sequentially by decomposing Eq.~\eqref{eq:truncexpon_joint} into a product of marginal and conditional probabilities
\begin{equation}
P_2(x_1,x_2) = P_2(x_1) P_2(x_2 \vert x_1),
\end{equation}
where 
\begin{equation}
    P_2(x_1) = \int_{x_1}^{b} \dd{x_2} P_2(x_1,x_2) = \frac{1-e^{-c(b-x_1)}}{c \mathcal{Z}_J}
\end{equation}
and $P_2(x_2 \vert x_1)$ is a one dimensional truncated exponential distribution in the variable $x_2 \in [x_1,b]$.

The CDF of the marginalized distribution is 
\begin{align}
    F_2(x_1) &= \int_{a}^{x_1} d x_1^\prime P_2(x_1^\prime) \nonumber \\ 
             & = \frac{1}{c^2 \mathcal{Z}_J} \left [ e^{-c(b-a)} - e^{-c(b-x_1)} - c(a-x_1) \right ].\label{eq:F2marginal}
\end{align}

Denote $y = F_2(x_1)$. To generate samples from the marginalized distribution $x_1 \sim P_2(x_1)$, one can sample $y \sim U(0,1)$ uniformly and then calculate $x_1$ by inverting $F_2$:
\begin{equation}
x_1 = F_2^{-1}(y).
\label{eq:x1_inv}
\end{equation}
An analytic solution for $x_1$ is possible in terms of the Lambert (product log) functions~\cite{Corless:1996co}, which are defined to invert the functional dependence $f(\alpha) = \alpha e^\alpha$. Since this map is not injective, its inverse
\begin{equation}
W_k(\alpha e^\alpha) = \alpha,\label{eq:W_alt}
\end{equation}
has multiple solution branches $k$. When $\alpha$ is real there are two solution branches; these are conventionally labeled $k=0$ for $\alpha \geq -1$, and $k=-1$ for $\alpha \leq -1$.

Now we will perform a series of algebraic manipulations on \Eqref{eq:x1_inv}. Begin by defining $B = -e^{-cb}$, and transform the dependent variable $y$ to a new one,
\begin{equation}
u = y c^2 Z_J + Be^{ca} + ca.\label{eq:udef}
\end{equation}
Referring to \Eqref{eq:F2marginal}, this yields the simplified constraint equation,
\begin{equation}
u = B e^{cx_1} + cx_1,\label{eq:udef2}
\end{equation}
or equivalently,
\begin{equation}
(u - cx_1) e^{(u-cx_1)} = B e^u.
\end{equation}
Next, apply $W_k$ to both sides and use \Eqref{eq:W_alt} with $\alpha = u - cx_1$.  The result is,
\begin{equation}
x_1 = \frac{1}{c} \left[u - W_k(B e^u) \right].\label{eq:x1_sol}
\end{equation}
Using \Eqref{eq:udef2}, we may write $\alpha = -\exp[-c(b-x_1)]$. Since $b - x_1 > 0$, the condition $\alpha \gtrless -1$ coincides with $c \gtrless 0$. It follows that we should select:
\begin{equation}
k = 0 \;\;\textrm{if}\;\; c \geq 0,\quad \textrm{or} \quad k = -1 \;\; \textrm{if}\;\; c \leq 0.
\end{equation}
Substitution of \Eqref{eq:udef} into the right of \Eqref{eq:x1_sol} then gives our final closed form solution for $x_1(y)$. Note that the limit $c \to 0$ is a removable singularity, for which $x_1 \to b - (b - a) \sqrt{1-y}$.

Our final procedure for sampling both $(x_1,x_2)$ from the joint distribution $P_2(x_1,x_2)$ can now be summarized as follows. Begin by generating a uniform random sample $y \sim U(0,1)$. Next, use the analytical solution of \Eqref{eq:x1_sol} with \Eqref{eq:udef} to generate a sample $x_1 \sim P_2(x_1)$, where $x_2$ has been marginalized out. Here, one may use an existing numerical subroutine to efficiently evaluate the Lambert function, $W_{0}$ or $W_{-1}$~\cite{Fukushima:2013rd}. With $x_1$ fixed, the second random variate, $x_2 \sim P_2(x_2 | x_1)$, can be directly sampled from the one dimensional truncated exponential distribution, \Eqref{eq:truncexpon_simple}, with the lower bound set to $a \to x_1$ and keeping the upper bound as $b$.



%

\begin{figure}[t]
\begin{center}
    \includegraphics[width=\figwidth]{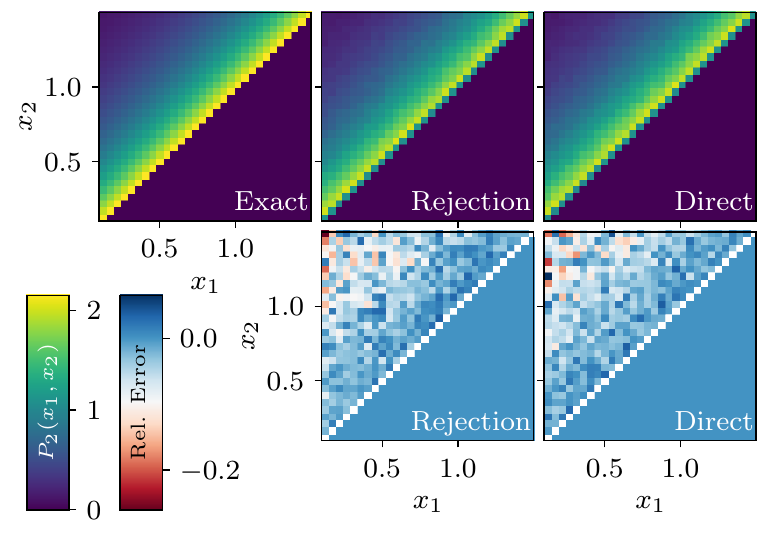}
\end{center}
\caption{Two dimensional truncated exponential distributions, \Eqref{eq:truncexpon_joint}, generated via rejection and direct sampling. The heatmaps on the top row show the exact, rejection sampled, and directly sampled distributions for parameters $a=0.1$, $b=1.5$ and $c=2.0$, with the relative error of each method compared to the exact shown in the bottom row. $8\times10^5$ samples of $x_1$ and $x_2$ each were used.}
\label{fig:joint_truncexpon_histogram_benchmark}
\end{figure}

\Figref{fig:joint_truncexpon_histogram_benchmark} shows results for samples drawn from the two dimensional probability distribution $P_2(x_1,x_2)$, \Eqref{eq:truncexpon_joint}, for a fixed set of parameters $a,b,$ and $c$. 
%
%
The leftmost heatmap shows the exact probability distribution $P_2(x_1,x_2)$ for comparison with the results obtained from rejection and direct sampling sampling for $8\times10^5$ random samples of $x_1$ and $x_2$ each. For a dataset of this size, both methods sample the exact distribution well. Looking at the relative error heatmaps corresponding to each sampling method, most regions are within $10\%$ of the exact distribution, with some areas in the top left having larger error due to the low sampling probability of this region. 

\begin{figure}[t]
\begin{center}
    \includegraphics[width=\figwidth]{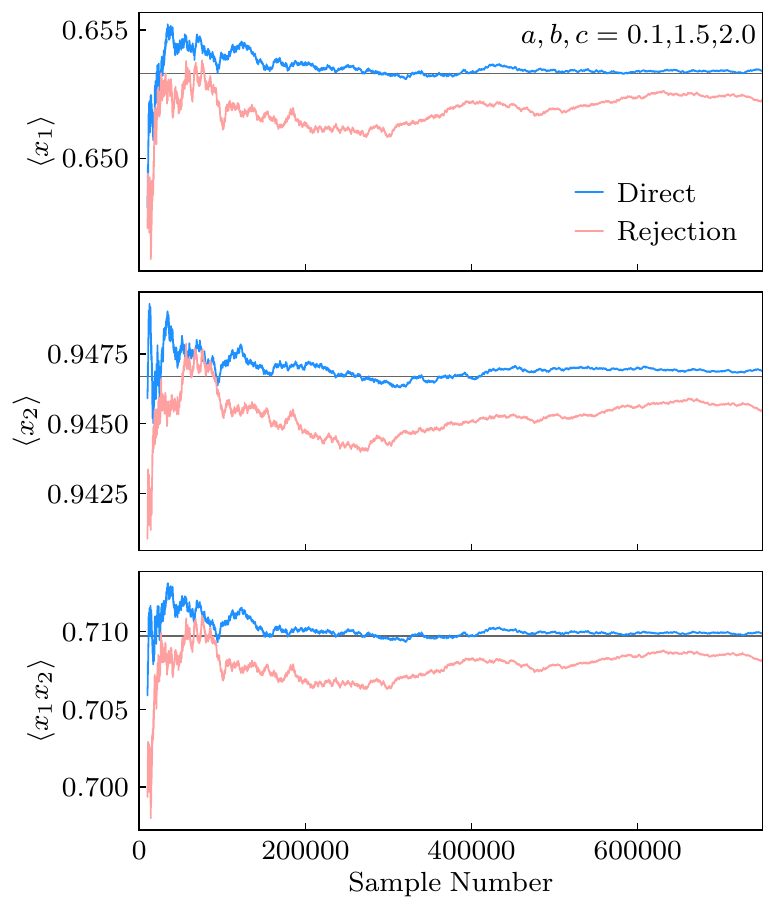}
\end{center}
\caption{Cumulative average of $\langle x_1 \rangle$, $\langle x_2 \rangle$, and $\langle x_1 x_2 \rangle$ as a function of number of samples. The values of $x_1$ and $x_2$ have been sampled from the two dimensional truncated exponential distribution, \Eqref{eq:truncexpon_joint}, via rejection sampling and direct sampling. For all three averages, the dataset obtained via direct sampling converges faster to the exact result (horizontal line).}
\label{fig:rvs_convergence}
\end{figure}

Due to the similarity between the rejection and direct sampling results in \Figref{fig:joint_truncexpon_histogram_benchmark}, it is not straightforward to determine directly which method is more efficient in reproducing the two dimensional truncated exponential distribution, \Eqref{eq:truncexpon_joint}. Kolmogorov-Smirnov tests in higher dimensions have been proposed~\cite{doi:10.1063/1.4822753,Justel:1997aa}, but technical issues make them highly non-trivial to implement, so we opt to compute running averages for the three quantities $\expval{x_1}$, $\expval{x_2}$, and $\expval{x_1 x_2}$ as a function of number of samples, with the results shown in \Figref{fig:rvs_convergence}. 
%
%
For all three quantities, the running average of the samples obtained via direct sampling converges faster to the exact values (obtained using \Eqref{eq:truncexpon_joint}, denoted by the horizontal line), than the rejection sampling dataset. 

In the next section, we show how sampling random variates from truncated exponential distributions, in both one and two dimensions, can improve the efficiency of some Quantum Monte Carlo simulations by reducing autocorrelation times between the samples.

\section{Application: Lattice Path Integral Quantum Monte Carlo}
\label{sec:qmc}

Markov chain Monte Carlo applications based on the Metropolis-Hastings algorithms create a Markov chain from configurations drawn according to a probability density function $\pi(\nu) = W(\nu)/\mathcal{Z}$, where $\nu$ denotes a configuration defined by the problem space.  Stochastic transition probabilities $T(\nu \to \nu^\prime)$ from a configuration $\nu$ to a new configuration $\nu^\prime$ should be independent of the history of the random walk. This is achieved via an ergodic set of Monte Carlo updates that satisfy the principle of detailed balance: $\pi(\nu) T(\nu \to \nu^\prime) = \pi(\nu^\prime) T(\nu^\prime \to \nu)$. Transition probabilities can be factored into a product of a selection probability $P(\nu \to \nu^\prime)$ and an acceptance probability $A(\nu \to \nu^\prime)$. From the principle of detailed balance, the acceptance ratio of a general Monte Carlo update can be expressed as:
\begin{equation}
\frac{A(\nu \to \nu^\prime)}{A(\nu^\prime \to \nu)} = \frac{W(\nu^\prime) P(\nu^\prime \to \nu)}{W(\nu) P(\nu \to \nu^\prime)} \equiv R.
\end{equation}
The $M$ configurations generated via the MCMC process can be utilized to approximate expectation values of observables: 
\begin{equation}
    \expval{O} = \sum_{\nu} O(\nu) \pi(\nu) \simeq \frac{1}{M} \sum_{i=1}^{M} O_i,
\label{eq:expvalGeneral}
\end{equation}
where $O_{i} = {O}(\nu_i)$. In practice, random samples $\nu \sim \pi(\nu)$ making
up the finite Markov chain $\qty{\nu_1,\dots, \nu_M}$ may not be independent,
leading to correlations in $O_i$ and $O_j$, i.e.\@ $\langle{O_iO_j}\rangle \ne
\langle O_i \rangle \langle {O_j} \rangle$. This can be quantified for observable $O$ via the integrated autocorrelation time $\mathcal{T}_O$:
\begin{equation}
    \mathcal{T}_O = 1 + 2 \sum_{\tau=1}^{M} \frac{\mathcal{C}_O(\tau)}{\mathcal{C}_O(0)},
\label{eq:TO}
\end{equation}
where the autocorrelation function is defined to be
\begin{equation}
    \mathcal{C}_O(\tau) =  \frac{1}{M-\tau}\sum_{i=1}^{M-\tau} 
    \qty(O_i - \expval{O}) \qty(O_{i+\tau}-\expval{O})\, .
\label{eq:correlation}
\end{equation}
Thus, truly independent measurements can only be performed for samples separated by $\mathcal{T}_O$ MCMC steps, and any algorithmic improvement leading to a reduction in $\mathcal{T}_O$ will improve the overall efficiency of a simulation.

In a recent work~\cite{CasianoDiaz:2022pf}, a subset of the authors of this paper introduced a path integral Monte Carlo algorithm for the simulation of bosonic lattice models at zero temperature ($T=0$), inspired by the finite temperature  PIMC Worm Algorithm~\cite{Prokofev:1998yf}.  We direct the reader to Ref.~\cite{CasianoDiaz:2022pf} for complete details of the algorithm which can be used to evaluate ground state expectation values: 
\begin{equation}
    \expval{O} \equiv \frac{\expval{O}{\Psi}}{\braket{\Psi}{\Psi}}
\label{eq:expval}
\end{equation}
by projection of a trial state $\vert \Psi_T \rangle$, with a large power of the density operator: $ \vert \Psi \rangle = \lim_{\beta\to\infty} e^{-\beta H} \vert \Psi_T \rangle$, where $H$ is the system Hamiltonian, and $\beta$ is the projection length.  

\begin{figure}[t]
\begin{center}
    \includegraphics[width=\figwidth]{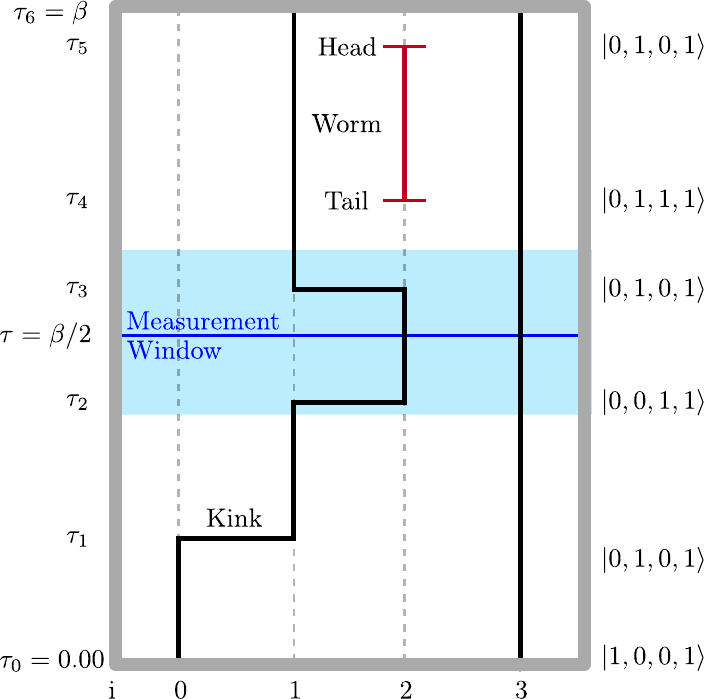}
\end{center}
\caption{Example of a worldline configuration in Path Integral Monte Carlo for 2 particles on 4 lattice sites. The paths propagate in the direction of imaginary time (vertical) and space (horizontal). Three kinks are shown at imaginary times $\tau_1,\tau_2,\tau_3$ and occur due to a particle hopping between adjacent sites. More configurations can be sampled by performing updates on the worm shown with tail at $\tau_4$ and head at $\tau_5$. Some estimators, like the kinetic energy, involve averaging a quantity, such as the number of kinks, over a window of user defined width centered around $\tau = \beta/2$.}
\label{fig:lattice_worldlines}
\end{figure}

Using the path integral formulation of quantum mechanics, the target configuration space can be represented as a set of paths, known as worldlines, that propagate in imaginary time (characterized by $\beta$) and space. \Figref{fig:lattice_worldlines} shows an example configuration of worldlines for a Bose-Hubbard model in one dimension with $N=2$ bosons on $L=4$ sites. 
%
%
The vertical direction represents imaginary time and the lattice sites span the horizontal direction. In this diagram, line widths are proportional to number of particles on a site, with dotted lines representing an empty site. The set of imaginary times $\lbrace \tau_i \rbrace$ labeled on the left side of the figure correspond to times at which the Fock state ($\ket{n_1,n_2,n_3,n_4}$) has changed (right side), where $n_j$ counts the number of particles on site $j$.  Local changes in occupation can occur via either kinks representing particle hops between adjacent sites, or via the insertion or deletion of a special type of truncated worldline known as a worm~\cite{Boninsegni:2006gc,Boninsegni:2006ed,Prokofev:1998nv,Prokofev:2001pu}. Formally, the worm tail and head represent bosonic creation and annihilation operators, respectively. The entire configuration space can be sampled by performing updates on the worm. 
 
The acceptance ratio for the insertion of a worm into the worldline configuration has the form:
\begin{equation}
R = \rm{const} \times \frac{1}{P(\tau_h,\tau_t)} \times e^{-c(\tau_h-\tau_t)} \;\;\;\;\; (a \leq \tau_t < \tau_h \leq b),
\label{eq:worm_insert_ratio}
\end{equation}
where $\tau_t$ and $\tau_h$ denote the imaginary times of the worm tail and head, respectively, and are randomly sampled from a joint probability distribution, $P(\tau_h,\tau_t)$. By choosing this distribution to be the two dimensional truncated exponential distribution, \Eqref{eq:truncexpon_joint}, the exponential factor in \Eqref{eq:worm_insert_ratio} cancels, leaving just a constant factor as the acceptance ratio:
\begin{equation}
R = \rm{const} \times \frac{1}{e^{-c(\tau_h-\tau_t)}} \times e^{-c(\tau_h-\tau_t)} = \rm{const},
\label{eq:worm_insert_ratio_simplified}
\end{equation}
where the normalization constant of the two dimensional truncated exponential from which $\tau_t,\tau_h$ are drawn has been absorbed into $\rm{const}$. We expect that this constant acceptance ratio can be further optimized by implementing a pre-equilibration stage that tunes simulation parameters via iterative methods, similarly to approaches for the tuning of the chemical potential, $\mu$, to set the average number of particles~\cite{PhysRevB.89.224502,PhysRevE.105.045311}. The acceptance ratios of the rest of the updates, which are related to either insertions and deletions of kinks or shifting worm ends in the imaginary direction, can also be reduced to a constant by sampling imaginary times from the one dimensional truncated exponential distribution, \Eqref{eq:truncexpon_simple}. For updates that shift worm ends in the imaginary time direction, sampling from this distribution actually leads to perfect direct sampling~\cite{Prokofev:1998yf,CasianoDiaz:2022pf} ($R = 1$).

In the discussion that follows, results for which imaginary times have been
sampled from a truncated exponential distribution, such that the acceptance
ratios take the form of \Eqref{eq:worm_insert_ratio_simplified}, will be
referred to as \emph{direct} sampling. The conventional \emph{rejection} scheme
instead involves sampling each of the imaginary times from a rescaled uniform
distribution, $\tau \sim {U}(a,b)$, where $a,b$ are the lower and upper bounds
of the interval. Thus, the joint probability
distribution for this case is $P(\tau_h, \tau_t) = 1/(b-a)^2$. However, both schemes still involve a Metropolis sampling step in which updates will be accepted by comparing if a random number, $r \sim {U}(0,1)$, satisfies $r < R$, and rejected otherwise. Formally, the sampled distribution is the same using both schemes, up to a pre-factor.

We benchmark the proposed direct sampling approach on a ground state quantum Monte Carlo simulation of the Bose-Hubbard model for itinerant bosons on a lattice~\cite{PhysRev.129.959}:
\begin{equation}
H = - t \sum_{ \langle i,j \rangle}  b_i^{\dag}  b_j^{\phantom{\dag}} + \frac{U}{2} \sum_i  n_i ( n_i-1) - \mu \sum_i  n_i\ ,
\label{eq:bh_hamiltonian}
\end{equation}
where $t$ is the tunneling between neighboring lattice sites $\langle i,j \rangle$, $U>0$ is a repulsive interaction potential, $\mu$ is the chemical potential, and $b_i^\dag(b_i^{\phantom{\dag}})$ are bosonic creation(annhilation) operators on site $i$, satisfying the commutation relation: $[{b}_i^{\phantom{\dag}},{b}_{j}^\dag]=\delta_{i,j}$, with $n_i = b_i^{\dag}b_i$ the local number operator. Simulations were performed in the canonical ensemble, in which $\mu$ is a simulation parameter. This model exhibits a quantum phase transition from a superfluid, at low interactions, to a Mott insulator, at strong repulsive interactions, where bosons become highly localized. The accurate determination of the quantum critical point has motivated much research~\cite{Carrasquilla:2013ya,PhysRevB.61.12474,PhysRevB.53.2691,PhysRevB.59.12184,PhysRevB.58.R14741,Kashurnikov:1996aa,PhysRevB.53.11776,Ejima_2011,doi:10.1063/1.3046265,PhysRevLett.96.180603,PhysRevB.44.9772,PhysRevLett.76.2937,PhysRevLett.65.1765,PhysRevA.86.023631,L_uchli_2008,PhysRevLett.108.116401} and, below, we report on simulations at a fixed interaction strength of $U/t = 3.3$, which is representative of the quantum critical regime where both spatial and temporal correlation lengths diverge, causing the well known problem of critical slowing down~\cite{Wolff:1990cs} where correlations between MCMC samples can be large. 

The kinetic energy estimator is non-diagonal in the Fock basis and is determined from the average number of kinks in the measurement window of \Figref{fig:lattice_worldlines}~\cite{CasianoDiaz:2022pf}:
\begin{equation}
    \langle \rm{K} \rangle = - \frac{\expval{N_{\rm{kinks}}}}{\Delta \beta}
\end{equation}
where $\Delta \beta$ is the window width. The potential energy estimator is diagonal in the Fock basis and is obtained by measuring 
\begin{equation}
    \expval{V} = \frac{U}{2} \sum_i \expval{n_i (n_i - 1)}
\end{equation}
at imaginary time $\tau = \beta/2$.

\begin{figure}[t]
\begin{center}
    \includegraphics[width=\figwidth]{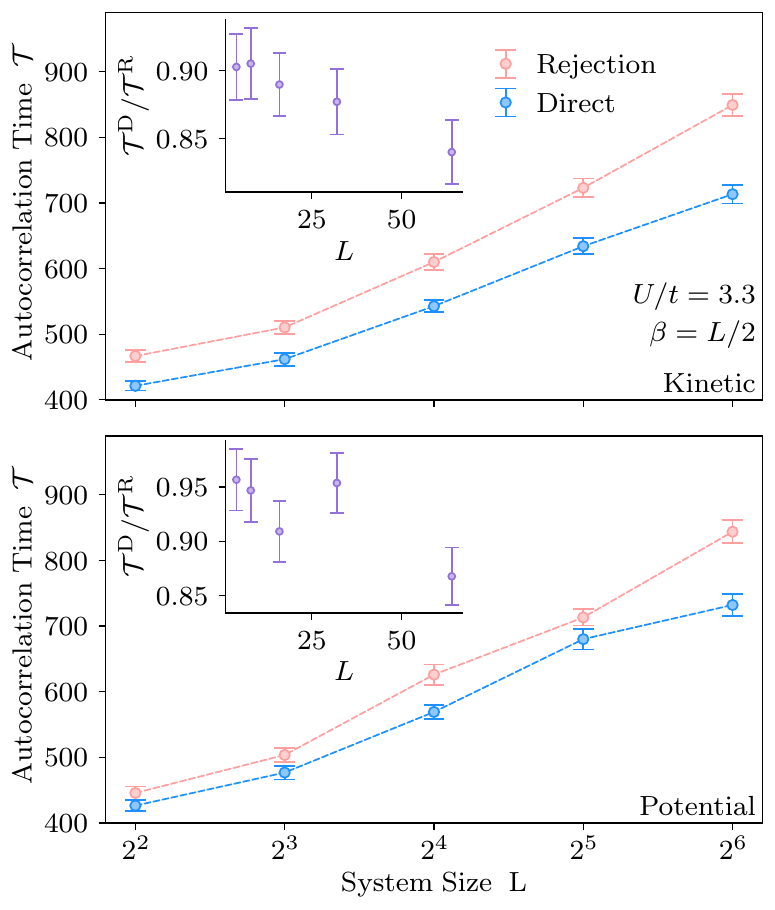}
\end{center}
\caption{Integrated autocorrelation times for the ground state kinetic (top)
    and potential (bottom) energy estimators of the one dimensional Bose-Hubbard model obtained via path integral Monte Carlo as a function of system size, $L$ at unit filing: $L=N$. Lower is better. The insets show the ratio of autocorrelation times obtained by sampling using direct and rejection (uniform distribution) methods for the one and two dimensional truncated exponential distributions. 
}
\label{fig:autocorr_vs_L}
\end{figure}

To understand the role of direct vs.\@ rejection sampling in our quantum Monte Carlo algorithm, we performed simulations of the one dimensional Bose-Hubbard model at unit-filling, $L=N$, with $L$ the number of sites and $N$ the number of particles, for different values of $L$ and computed the integrated autocorrelation time in Eq.~\eqref{eq:TO} using both sampling methods at $U/t = 3.3$, near the superfluid-insulating critical point. The results are shown in \Figref{fig:autocorr_vs_L} where the autocorrelation times were computed using the $\rm{emcee}$ Python library~\cite{Foreman-Mackey_2013}, which is based on the methods described in Refs.~\cite{2010CAMCS...5...65G,Sokal1996MonteCM}.
%
%
For system sizes up to $L = 2^6 = 64$, the autocorrelation time for both the kinetic and potential energy was lower for the case in which imaginary times were directly sampled from truncated exponential distributions when performing worm updates. The insets show the ratio of autocorrelation times sampling from truncated exponential distributions (direct) over autocorrelation times sampling from uniform distributions (rejection), $\mathcal{T}^{\rm{D}}/\mathcal{T}^{\rm{R}}$, as a function of system size. A ratio less than unity indicates a decrease in correlations amongst samples, and improvements of $~\sim 15\%$ are observed for the largest system sizes studied. 

Direct sampling has a larger effect on the autocorrelation time of the kinetic energy estimator as the simulation dynamics of the average number of kinks is directly related to improved worm dynamics in the simulation (see \Figref{fig:lattice_worldlines}).
Error bars represent the standard error of the mean autocorrelation time computed from $80$ independent simulations.

Since the lattice ground state PIMC algorithm described in Ref.~\cite{CasianoDiaz:2022pf} is a projection algorithm, it is subject to a systematic error that decreases with increasing projection length, $\beta$. Due to this, reliable estimates of observables are obtained by performing simulations for various values of $\beta$, and extrapolating the exact value, within error bars,  from an exponential fit in $\beta$ plus an additive constant: $ \expval{{O}(\beta)} = C_1 e^{-\beta C_2} + \expval{O}$, where $C_1,C_2,$ and $\expval{O}$ are fitting parameters. Thus, to understand the role of direct sampling on the $\beta$ extrapolation, we plot $\beta$ dependent autocorrelation times for the kinetic and potential energies for a fixed system size $L = 12$ and $U/t = 3.3$ in \Figref{fig:autocorr_vs_L}.
%
\begin{figure}[t]
\begin{center}
    \includegraphics[width=\figwidth]{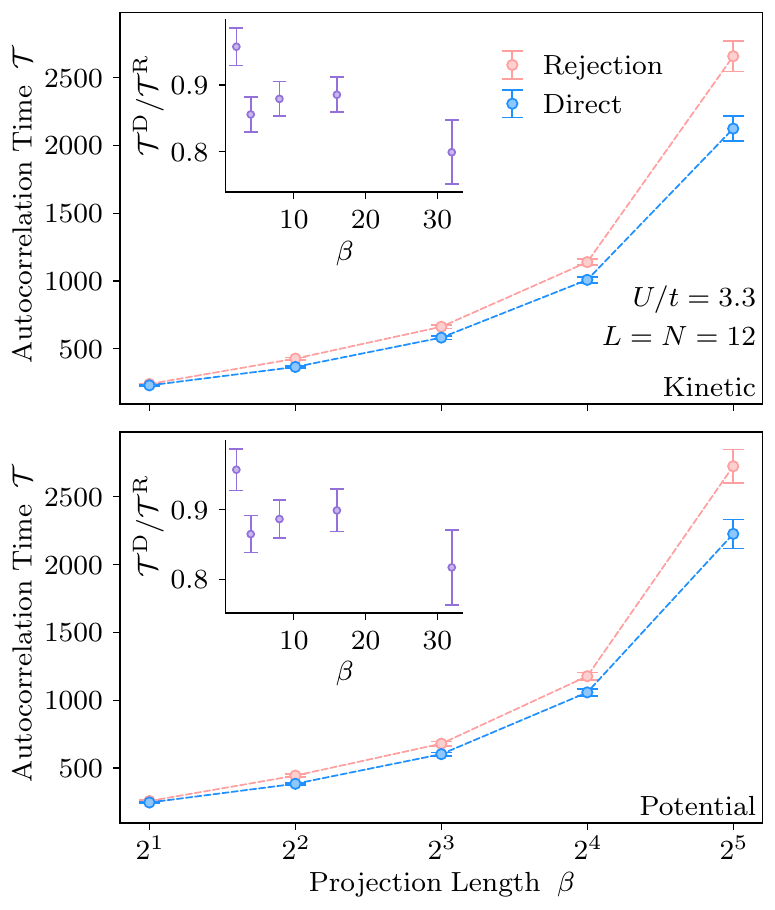}
\end{center}
\caption{Autocorrelation times for the ground state kinetic (top) and potential
    (bottom) energy estimators of the one dimensional Bose-Hubbard Model obtained via Path Integral Monte Carlo as a function of projection length, $\beta$.  Lower is better. 
The insets show the ratio of autocorrelation times obtained by sampling using direct and rejection (uniform distribution) methods for the one and two dimensional truncated exponential distributions.} 
\label{fig:autocorr_vs_beta}
\end{figure}
%
The autocorrelation times are once again seen to improve by sampling imaginary times directly from truncated exponential distributions, with the insets showing time reductions of approximately $20\%$ for the largest $\beta$ values.

%
\begin{figure}[t]
\begin{center}
    \includegraphics[width=\figwidth]{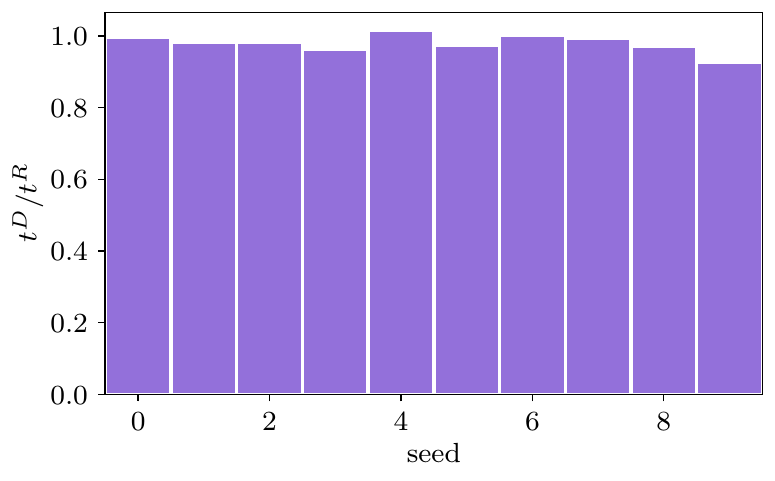}
\end{center}
\caption{Wall clock times for path integral Monte Carlo simulations for a $1D$ Bose-Hubbard lattice of size $L=N=12$ at $U/t=3.3$ using rejection (R) and direct (D) sampling of the truncated exponential distribution. Results are shown for $10$ independent runs for each method starting from different random seeds and $10^6$ samples of the kinetic and potential energies were collected. The wall times did not change significantly when using different sampling schemes, with an average ratio of wall times $t^{\rm{D}}/t^{\rm{R}} = 1.02 +/- 0.01$.}
\label{fig:wall_times}
\end{figure}
%
The above results demonstrate that sampling directly from truncated exponential distributions results in decreased autocorrelation times for estimators in the algorithm presented in~\cite{CasianoDiaz:2022pf}. However, for our implementation of the algorithm, it was also seen that wall clock times in the direct sampling of truncated exponential distributions were no slower than the original version, where imaginary times where sampled from uniform distributions.  In other words, the direct sampling scheme can be implemented without impacting practical run times.  For a system of $L=N=12$ bosons at $U/t=3.3$ and $\beta=16$, we performed $10$ simulations, each starting from different random seeds, and observe that the fraction of wall clock times using both sampling schemes was $t^{\rm{D}}/t^{\rm{R}} = 1.02 +/- 0.01$.  The fraction of wall times for each seed, for the direct over rejection methods, $t^{\rm{D}}/t^{\rm{R}}$, are shown in \Figref{fig:wall_times}. 
The new sampling scheme has thus successfully reduced autocorrelation times without slower wall times, resulting in an effective speedup of the quantum Monte Carlo application discussed.

All code, scripts and data needed to confirm the results presented in this section are available in open source repositories~\cite{codeRepoTruncExpon,zenodo,repo}. 

\section{Conclusions}
\label{sec:conclusion}

In this paper, we have shown how to directly obtain random variates from two dimensional truncated exponential distributions via a two step inverse sampling method. The dataset of random variates obtained directly from truncated exponential distributions, in both one and two dimensions, better reproduced the target distribution for a finite number of samples. Direct sampling of the truncated exponential distribution was then applied to lattice path integral Monte Carlo, where this distribution appears in the acceptance probability of \emph{worm} updates and enabled the efficient sampling of the imaginary time worldline configuration space.  
The direct sampling approach leads to reduced integrated autocorrelation times for the kinetic and potential energy estimators, while keeping the simulation wall times practically the same between both sampling schemes.  For the system sizes considered, overall efficiency gains of $15\%$ are identified.

Future avenues for research include implementing iterative methods to optimize non-physical algorithmic parameters that appear in the now constant worm update acceptance ratio to further improve dynamics and approach the ideal sampling limit.

\section{Acknowledgements}

We thank N. Prokof’ev and M. Thamm for fruitful discussions. 

This work was partly supported by the Laboratory Directed Research and Development funding of Los Alamos National Laboratory (LANL). LANL is operated by Triad National Security, LLC, for the National Nuclear Security Administration of U.S. Department of Energy (Contract No. 89233218CNA000001). K.~B., Y.~W.~L., and A.~D. acknowledge support by the U.S. Department of Energy, Office of Science, Office of Basic Energy Sciences, under Award Number DE-SC0022311.


\bibliographystyle{elsarticle-num}
\bibliography{refs}

\end{document}